\documentclass[
aps,
reprint,
superscriptaddress,
]{revtex4-2}

\usepackage[utf8]{inputenc}
\usepackage[T1]{fontenc}

\usepackage{graphicx}   
\usepackage{dcolumn}    
\usepackage{booktabs}   
\usepackage{multirow}   
\usepackage{bm}         

\usepackage{amsmath,amssymb,amsfonts}  
\usepackage{mathrsfs}                  
\DeclareMathAlphabet{\mathcal}{OMS}{cmsy}{m}{n}
\DeclareSymbolFont{largesymbols}{OMX}{cmex}{m}{n}

\usepackage{siunitx}
\usepackage[version=4]{mhchem}

\usepackage[
colorlinks=true,
pdfstartview=FitV,
linkcolor=blue,
citecolor=blue,
urlcolor=blue
]{hyperref}

\begin{document}
	
	\title{Magnetic Skyrmion Interacting with Optical Skyrmion}
	
	\author{Lan Bo}
	\thanks{These authors contributed equally}
	\affiliation{Department of Applied Physics, Waseda University, Okubo, Shinjuku-ku, Tokyo 169-8555, Japan}
	\author{Jian Chen}
	\thanks{These authors contributed equally}
	\affiliation{School of Optical-Electrical and Computer Engineering, University of Shanghai for Science and Technology, Shanghai 200093, China}
	\author{Xichao Zhang}
	\affiliation{Department of Applied Physics, Waseda University, Okubo, Shinjuku-ku, Tokyo 169-8555, Japan}
	\affiliation{Department of Electronic and Computer Engineering, The Hong Kong University of Science and Technology, Clear Water Bay, Kowloon, Hong Kong, China}
	\affiliation{Department of Physics, The Hong Kong University of Science and Technology, Clear Water Bay, Kowloon, Hong Kong, China}
	\affiliation{IAS Center for Quantum Matter, The Hong Kong University of Science and Technology, Clear Water Bay, Kowloon, Hong Kong, China}
	\author{Yan Zhou}
	\email[Corresponding E-mail: ]{zhouyan@cuhk.edu.cn}
	\affiliation{School of Science and Engineering, The Chinese University of Hong Kong, Shenzhen, Guangdong 518172, China}
	\author{Chengwei Qiu}
	\email[Corresponding E-mail: ]{chengwei.qiu@nus.edu.sg}
	\affiliation{Department of Electrical and Computer Engineering, National University of Singapore, Singapore 117583, Singapore}
	\author{Masahito Mochizuki}
	\email[Corresponding E-mail: ]{masa_mochizuki@waseda.jp}
	\affiliation{Department of Applied Physics, Waseda University, Okubo, Shinjuku-ku, Tokyo 169-8555, Japan}

	\begin{abstract}
		Magnetic skyrmions (MSks) and optical skyrmions (OSks) embody topology in matter and in light, respectively.
		Here we investigate the interaction between a single MSk and an OSk beam. Three distinct nonlinear dynamical modes are identified: rotation, skipping, and trochoidal motion. 
		By decomposing the optical driving force into gradient, orbital–angular–momentum, and spin–angular–momentum contributions, we clarify their respective roles of radial confinement, azimuthal drift, and precessional modulation. 
		The skipping motion arises from the azimuthal asymmetry of the OSk beam and exhibits spatial selectivity originating from the magnetization-polarization coupling between the MSk and OSk.
		In three dimensions, the coupling acquires a propagation-dependent phase dominated by the differential Gouy phase, which yields $z$-asymmetric skipping trajectories.
		These results bridge topological particles and topological fields within a unified framework, offering helicity-selective and phase-programmable routes to optomagnonic control.
		
	\end{abstract}

	\maketitle
	
	\textit{Introduction.}--
	Topology has emerged as a universal organizing principle in modern physics, governing phenomena across scales from the cosmological to the quantum\cite{volovik2003universe}. 
	Among its most celebrated manifestations is the \textit{skyrmion}, a localized field configuration first conceived in nuclear field theory\cite{skyrme1962unified}. 
	Since then, skyrmions have displayed remarkable ubiquity, appearing in quantum Hall states\cite{sondhi1993skyrmions}, Bose–Einstein condensates\cite{al2001skyrmions}, and superfluids\cite{donati2016twist}, as well as in magnetic\cite{muhlbauer2009skyrmion}, ferroelectric\cite{das2019observation}, and liquid-crystal materials\cite{smalyukh2010three,zhang2021autonomous,asilehan2025light,asilehan2025monopole}. Their presence has recently expanded to wave-based systems including plasmonic\cite{tsesses2018optical}, photonic\cite{gao2020paraxial}, acoustic\cite{ge2021observation}, and water-wave media\cite{smirnova2024water}. 
	The cross-disciplinary presence highlights skyrmions as a unifying framework for geometry\cite{manton1987geometry},
	fractional statistics\cite{wilczek1982quantum}, and particle-like 
	dynamics\cite{jiang2017direct}.
	
	Magnetic skyrmions (MSks) have become the most extensively studied owing to their direct real-space observation\cite{yu2010real} and high controllability\cite{iwasaki2013current}. 
	They can be stabilized either in noncentrosymmetric magnets, where bulk Dzyaloshinskii–Moriya interactions (DMI) favor Bloch textures\cite{muhlbauer2009skyrmion}, or in thin films, where interfacial DMI leads to Néel textures\cite{sampaio2013nucleation}. 
	Their current-driven dynamics manifest emergent electrodynamics, most notably the skyrmion Hall effect\cite{jiang2017direct}, revealing the intrinsic
	link between topology and charge transport. 
	Owing to their robustness and tunability, MSks constitute a versatile platform for topological spintronics, with applications in nonvolatile memories\cite{fert2013skyrmions}, logic devices\cite{zhang2015magnetic,zhang2025nanofluidic}, microwave technologies\cite{mochizuki2012spin}, and neuromorphic computing\cite{huang2017magnetic}.
	
	The recognition of MSks has, in turn, inspired their optical counterparts—optical skyrmions (OSks)\cite{shen2024optical}. 
	OSks were first reported in the plasmonic near field of surface plasmon polaritons\cite{tsesses2018optical} and were subsequently extended to diverse vector-field platforms, from real-space electric and magnetic field distributions\cite{davis2020ultrafast,shen2021supertoroidal,deng2022observation} to artificial spaces defined by photonic spin or Stokes vectors\cite{du2019deep,dai2020plasmonic,sugic2021particle,shen2022generation}.
	Various coupling mechanisms have been exploited to engineer OSks, including spin–orbit interaction\cite{du2019deep,dai2020plasmonic}, photonic quantum spin Hall effect\cite{zhang2021bloch}, and artificial gauge fields\cite{krol2021observation}. 
	Furthermore, OSks realized in free-space structured light offer exceptional flexibility and a rich diversity of topological configurations\cite{sugic2021particle,shen2022generation}, opening new opportunities for precision metrology\cite{yang2023spin}, optical manipulation\cite{wang2024topological}, and information processing\cite{wang2025perturbation,ma2025nanophotonic}.
	
	MSks and OSks represent two widely studied realizations of topology in matter and in light.
	Although both are characterized by the same topological charge $Q = (4\pi)^{-1}\!\int \mathbf{n}\!\cdot (\partial_x\mathbf{n}\!\times\!\partial_y\mathbf{n})\,\mathrm{d}x\,\mathrm{d}y$, with $\mathbf{n}$ specifying the local orientation, they have been investigated almost exclusively within their respective domains\cite{chen2025more}.
	This separation leaves open the fundamental question of how the spin topology of magnetic order interacts with the field topology of light.
	Here, we study how a single MSk quasiparticle responds to an OSk beam.
	Unlike conventional optical drives that are predominantly linear or rotationally symmetric, OSks exploit the orbital and spin angular momentum (OAM and SAM) of structured light to realize a nontrivial Stokes-vector topology, which can remain well defined under field distortions and is thus favorable for precision control.
	Moreover, the coherent superposition of two OAM modes generates a controllable azimuthal modulation that breaks rotational symmetry and introduces spatial selectivity.
	We identify three nonlinear dynamical modes—rotation, skipping, and trochoidal motion—
	establishing a topology-enabled optical handle for contact-free and programmable MSk manipulation.

\begin{figure}[t]
		\includegraphics[width=0.5\textwidth]{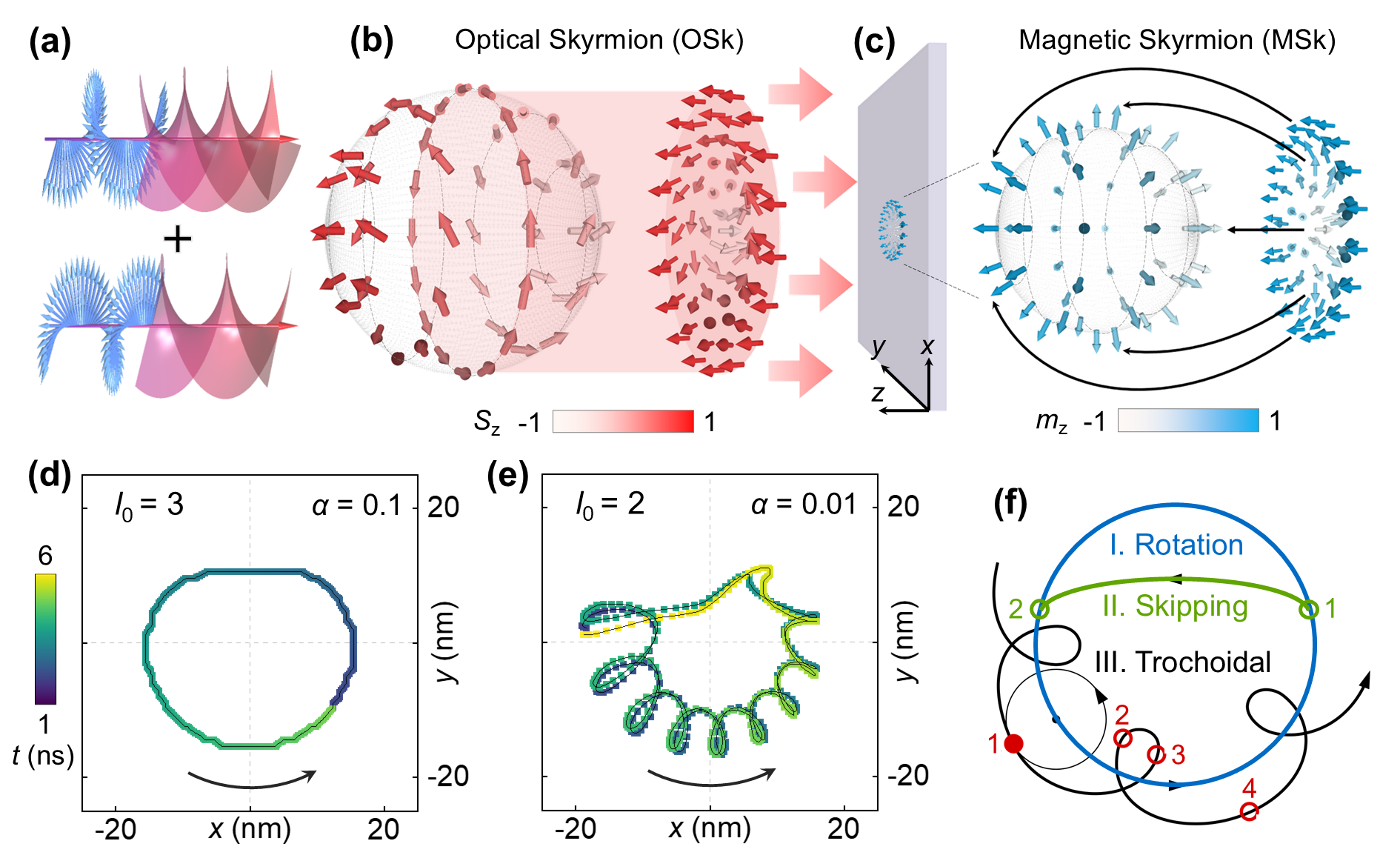}
		\caption{
			(a) 
			Two orthogonally polarized LG modes with $l_0=2$ and $l_1=3$. 
			Red helical surfaces and blue arrows represent the OAM and SAM.
			(b),(c) 
			Bloch-type OSk and Néel-type MSk shown through their stereographic projections.
			(d),(e) 
			Representative MSk trajectories under different $\alpha$ and $l_0$.
			(f) 
			Schematic summary of the three dynamical modes.
		}
		\label{fig:1}
	\end{figure}

	\textit{Model of MSk–OSk Interaction.}--
	The OSk beam is constructed by superposing two Laguerre–Gaussian (LG) modes,
	$\mathbf{\Psi}(\mathbf r)=u_0(\mathbf r)\,\mathbf e_L + u_1(\mathbf r)\,\mathbf e_R e^{i\theta_0}$,
	where $\theta_0$ is the phase difference,
	$\mathbf e_{L,R}=\tfrac{1}{\sqrt{2}}(1, \pm i)^{\rm T}$
	denote the left- and right-handed circular polarization bases. The LG modes
	\begin{equation}
		u_{n,l}(\rho,\phi) = B_0\!\left(\frac{\rho}{W_0}\right)^{|l|}
		e^{-\rho^2/W_0^2}
		L_n^{|l|}\!\!\left(\frac{2\rho^2}{W_0^2}\right)
		e^{i l\phi},
		\label{eq:LG2D}
	\end{equation}
	are	defined at the focal plane $z=0$.
	Here, $B_0$, $W_0$, $n$, $l$, and $L_n^{|l|}$ denote the beam amplitude, waist, radial index, azimuthal index, and Laguerre polynomial, respectively.
	Figure~\ref{fig:1}(a) illustrates two LG modes $u_0\mathbf e_L$ and $u_1\mathbf e_R$ with $l_0=2$ and $l_1=3$: the OAM specified by $l$ appears as red helical wavefronts, while the circular bases $\mathbf e_{L,R}$ encode the SAM shown by blue arrows. 
	The local field is conveniently described by normalized Jones spinor 
	$|\psi(\mathbf r)\rangle = (\mathbf e_L + \eta(\mathbf r)\,\mathbf e_R)/\sqrt{1+|\eta(\mathbf r)|^2}$,
	with complex mode ratio $\eta(\mathbf r)=e^{i\theta_0}u_1(\mathbf r)/u_0(\mathbf r)$.
	The corresponding Stokes vector
	$\mathbf S(\mathbf r)=\langle\psi(\mathbf r)|\boldsymbol\sigma|\psi(\mathbf r)\rangle$
	maps the plane to the Poincaré sphere and realizes a skyrmion texture when $n_0=n_1=0$ and $l_1=l_0+1$, yielding  $Q\equiv l_1-l_0=1$, as schematically shown in Fig.~\ref{fig:1}(b).
	
	We adopt an optomagnetic effective-field picture in which the structured optical field is treated as an externally prescribed drive~\cite{fujita2017encoding,yang2018photonic,jiang2020twisted,guan2023optically,lei2024rotational,zhang2026skyrmion}, neglecting magnetization-to-optical feedback. 
	This framework isolates coherent effective-field driving, while a distinct photothermal route in which heating and thermal gradients contribute to MSk motion has been considered elsewhere~\cite{fujita2017ultrafast}.
	The validity conditions of this coherent-drive description and its robustness against thermal perturbations are assessed in the Supplemental Material\cite{SM}.
	\nocite{vansteenkiste2014design,heeres2014subwavelength}
	The above optically induced field is given by
	\begin{equation}
		\mathbf B_{\mathrm{opt}}(\mathbf r,t)
		= \mathrm{Re}[\mathbf\Psi(\mathbf r)]\cos(\omega_0 t)
		+ \mathrm{Im}[\mathbf\Psi(\mathbf r)]\sin(\omega_0 t),
	\end{equation}
	where $f=\omega_0/2\pi$ denotes the envelope modulation frequency. 
	This complex field drives the MSk in the magnetic layer, as shown in Fig.~\ref{fig:1}(c), whose dynamics obey the modified Landau–Lifshitz–Gilbert (LLG) equation,
	\begin{equation}
		\partial_t \mathbf m = -\gamma\, \mathbf m \times \left( \mathbf B_{\mathrm{int}} + \mathbf B_{\mathrm{opt}} \right) + \alpha\, \mathbf m \times \partial_t \mathbf m,
	\end{equation}
	where $\mathbf m$, $\gamma$, and $\alpha$ are the magnetization vector, gyromagnetic ratio, and Gilbert damping constant, respectively. 
	The internal field $\mathbf B_{\mathrm{int}}$ is derived from the micromagnetic energy density, and its explicit expression, along with all optical and magnetic parameters used in the simulations, is detailed in the Supplemental Material\cite{SM}.
	
	\textit{Overview of Three Dynamical Modes.}--
	We investigate the dynamics of a single N\'eel MSk driven by a Bloch OSk.
	The MSk core trajectories $\mathbf R(t)$, defined as the position of the minimum $m_z$, are shown in Figs.~\ref{fig:1}(d) and ~\ref{fig:1}(e).
	Representative cases $(\alpha,l_0)=(0.1,3)$ and $(0.01,2)$ both exhibit counterclockwise motion. 
	The full parameter set and comparisons with alternative $\mathbf R(t)$ definitions are provided in the Supplemental Material\cite{SM}.
	The trajectories display three characteristic dynamical components: an underlying rotation, localized skipping event, and an additional trochoidal modulation, as schematically shown in Fig.~\ref{fig:1}(f).
	The rotation and skipping coexist within the same trajectory, while the trochoidal mode develops continuously as the damping is reduced.  
	At higher damping [Fig.~\ref{fig:1}(d)], the MSk undergoes nearly circular rotation interrupted by localized radial excursions, giving rise to skipping.
	At lower damping [Fig.~\ref{fig:1}(e)], the trajectory develops into a looped structure superimposed on a net drift.
	As shown by the black curve in Fig.~\ref{fig:1}(f), such a trajectory is termed an \emph{epicycle}\cite{evans2013mechanical}, originally introduced in astronomy, whereas in the context of magnetization dynamics we adopt the term trochoidal\cite{ritzmann2018trochoidal} in what follows.
	
\begin{figure}[t]
		\includegraphics[width=0.5\textwidth]{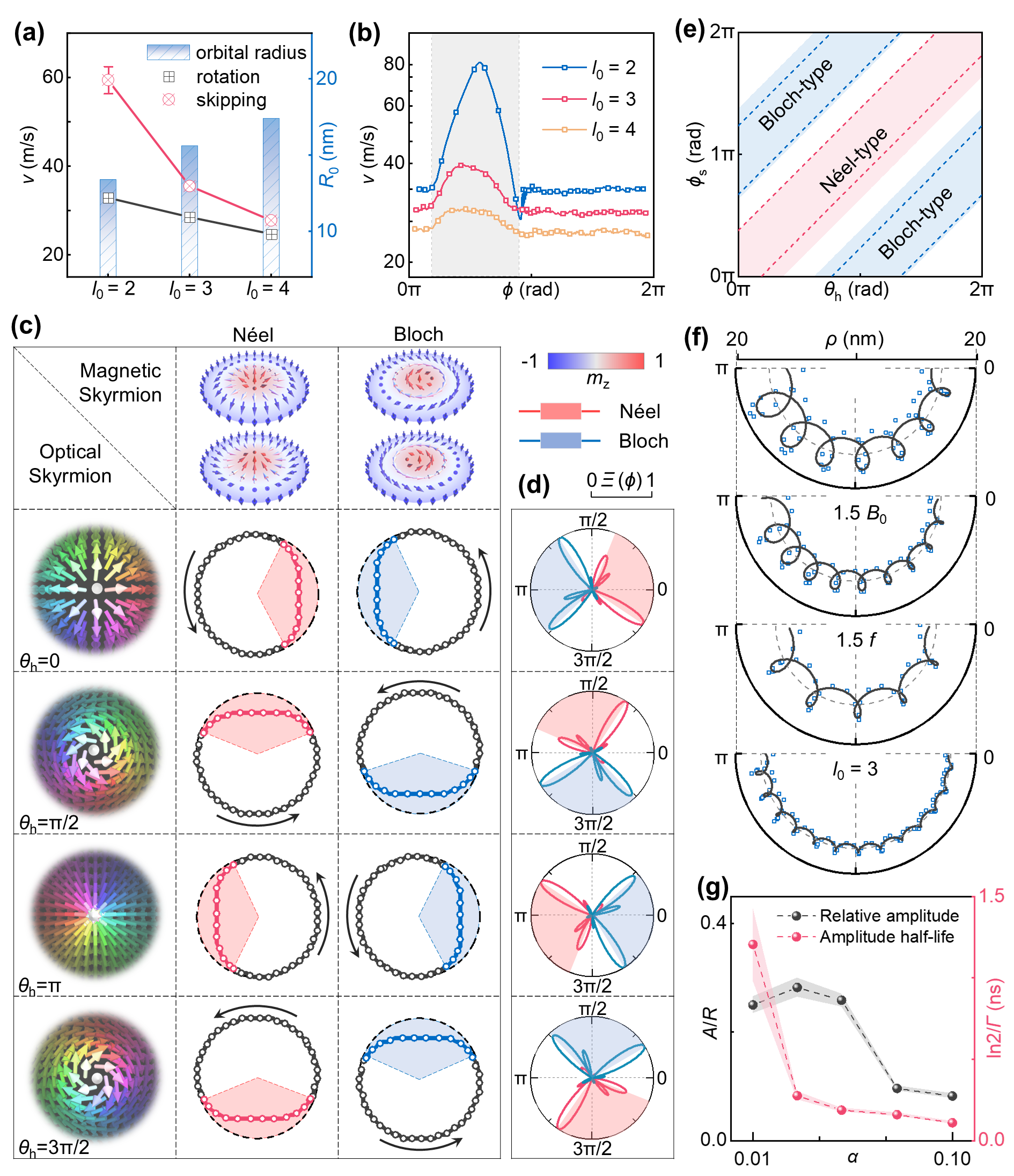}
		\caption{
			(a)
			Rotation/skipping velocities and the orbital radius $R_0$ versus $l_0$.
			(b)
			Azimuthal dependence of the instantaneous velocity $v(\phi)$ for different $l_0$. The shaded bands mark the skipping window.
			(c)
			Trajectories for different types of OSks and MSks. 
			(d)
			Polar plots of the skipping index $\varXi(\phi)$ for Néel/Bloch MSks, with predicted skipping sectors shaded.
			(e)
			Skipping azimuth $\phi_\mathrm{s}$ from simulations (shaded bands) compared with analytical predictions (dashed lines).
			(f)
			Simulated trochoidal trajectories (symbols) and corresponding fits (curves) for varying $B_0$, $f$, and $l_0$.	
			The top panel reproduces the reference case in Fig.~\ref{fig:1} (e).
			(g)
			Relative amplitude $A/R$ and amplitude half-life $\ln2/\varGamma$ versus $\alpha$.
		}
		\label{fig:2}
	\end{figure}

	\textit{Rotation and Skipping Motion.}--
	We first focus on the rotation and skipping motions under $\alpha = 0.1$.
	Figure~\ref{fig:2}(a) shows the orbital radius $R_0$ and the average velocities as functions of $l_0$.
	With increasing $l_0$, the intensity maximum of the OSk shifts outward, leading to a larger $R_0$, while the velocity decreases due to the weakened radial field gradient at larger radii.
	The skipping velocity remains higher than the rotational one, reflecting transient inward acceleration. 
	The difference between the two velocities diminishes for larger $l_0$, because the differential term $\Delta l$ contributes less strongly, consistent with the increased velocity fluctuations indicated by the error bars. 
	Figure~\ref{fig:2}(b) plots the instantaneous velocity $v(\phi)$ versus the azimuthal angle $\phi$. 
	Uniform rotation yields an almost constant $v$, whereas sharp variations appear only within the skipping sectors.
	For fixed $\theta_0=\pi/2$, the skipping window (shaded regions) is almost independent of $l_0$.
	
	To locate the skipping and study its spatial selectivity, we varied $\theta_0$ (which directly sets the OSk helicity $\theta_{\mathrm{h}}$) and considered both N\'eel- and Bloch-type MSks.
	Figure~\ref{fig:2}(c) summarizes the resulting trajectories for different OSk–MSk combinations.
	The shaded wedges mark the skipping, characterized by the skipping azimuth $\phi_{\mathrm{s}}$, which varies nearly linearly with $\theta_h$, indicating that the skipping direction is locked to the OSk helicity.
	In contrast, N\'eel and Bloch MSks exhibit a pronounced two-fold symmetry, with their skipping windows shifted by $\pi$.
	These trends are accounted for by an analytical model based on the optical Zeeman interaction (see Supplemental Material\cite{SM}).
	A normalized skipping index $\varXi(\phi)$ is introduced to quantify the skipping, whose two peaks identify the onset and recovery of skipping, as shown in Fig.~\ref{fig:2}(d).
	The extracted $\phi_{\mathrm{s}}$ (shaded bands) for different $\theta_\mathrm{h}$ are summarized in Fig.~\ref{fig:2}(e), together with the analytical curves (dashed lines).
	The theoretical predictions agree well with the simulated trajectories, yielding two parameter-free relations:
	(i) $\phi_{\mathrm{s}}^{B} = \phi_{\mathrm{s}}^{N} + \pi$, and
	(ii) $\phi_{\mathrm{s}}(\theta_0 + \Delta\theta_0)=\phi_{\mathrm{s}}(\theta_0) + \Delta\theta_0$.
		Together, these results establish OSk-helicity-locked and MSk-chirality-shifted trajectory selection, pointing to the phase-programmable magnetic-signal transduction scheme demonstrated in the Supplemental Material\cite{SM}.
	
	\textit{Trochoidal Motion.}--
	Motivated by the epicyclic geometry and magnetic damping, we model the trochoidal motion using a damped epicyclic form,
	\begin{equation}
		\begin{aligned}
			x(t) &= R\cos(\varOmega t) + A e^{-\varGamma t}\cos(\omega t), \\
			y(t) &= R\sin(\varOmega t) + A e^{-\varGamma t}\sin(\omega t),
		\end{aligned}
		\label{eq:trochoid_fit}
	\end{equation}
	where $R$, $A$, $\varGamma$, $\varOmega$, and $\omega$ are fitting parameters 
	representing the base-orbit radius, oscillation amplitude, damping rate, and the orbital and oscillation frequencies, respectively. 
	To exclude skipping, only the lower half of the orbit is used for fitting, with $t$ defined locally.
	
	Figure~\ref{fig:2}(f) compares simulated trochoidal trajectories with fits to Eq.~(\ref{eq:trochoid_fit}), showing excellent agreement.
	The top panel corresponds to the reference case in Fig.~\ref{fig:1}(e).
	In the small-amplitude limit, the instantaneous radius is approximated by
	$R + A e^{-\varGamma t}\cos[(\omega-\varOmega)t]$,
	where the detuning $\omega-\varOmega$ sets the number of loops.
	The lower panels show trajectories obtained by increasing $B_0$, $f$, and $l_0$ by a factor of 1.5.
	A larger $B_0$ reduces $A$ and increases $|\omega|$, indicating tighter radial confinement and denser loops.
	Increasing $f$ enhances the orbital frequency $\varOmega$ while slightly reducing $R$ and $A$, producing a more compact and faster orbital rotation.
	By contrast, increasing $l_0$ expands the orbit and flattens the radial potential, leading to reduced $\varOmega$ and $A$ with a weakly enhanced $|\omega|$.
	Together, these trends demonstrate tunable trochoidal dynamics via the field strength, temporal modulation, and spatial structure of the OSk beam.

	In addition to the OSk parameters, we also examined the role of $\alpha$, which controls the energy dissipation of the MSk motion.
	Figure~\ref{fig:2}(g) shows the evolution of the trochoidal dynamics with increasing $\alpha$.
	Both the relative amplitude $A/R$ (left axis) and the amplitude half-life $(\ln 2)/\varGamma$ (right axis) decrease monotonically, indicating that stronger damping suppresses radial excursions and accelerates the decay of the oscillatory envelope.
	As dissipation increases, the trochoidal motion gradually collapses into a purely circular orbit, marking the transition from the underdamped to the overdamped regime.

	\textit{Optical forces and field topology.}--
	The MSk center dynamics under optical driving are described by the Thiele equation,
	\begin{equation}
		\mathbf G\times \mathbf v
		+ \alpha\,\mathcal{D}\,\mathbf v
		= \mathbf F_{\rm grad}
		+ \mathbf F_{\rm OAM}
		+ \mathbf F_{\rm SAM},
		\label{eq:thiele}
	\end{equation}
	where $\mathbf G = G\,\hat{\mathbf z}$ with $ G = -4\pi Q $ is the gyrotropic vector, and $\mathcal{D}$ is the dissipative tensor.
	Within the rigid-particle approximation (see Supplemental Material\cite{SM} for its validity and limitations), Eq.~(\ref{eq:thiele}) balances the Magnus and dissipative forces against three optical driving forces.
	The optical forces arise respectively from the intensity gradient, the OAM-associated azimuthal phase winding, and the spin–curl effect due to the spatial variation of local SAM. 
	Their explicit forms are\cite{yang2025optical}
	\begin{align}
		\mathbf F_{\rm grad}
		&= -\,\kappa_{\rm g}/(4\mu_0)\;
		\nabla\!\left(|B_x|^2 + |B_y|^2\right),
		\label{eq:Fgrad}\\
		\mathbf F_{\rm OAM}
		&= \kappa_{\rm O}/\mu_0\;
		\mathrm{Im}\!\left(
		B_x^\ast\nabla B_x + B_y^\ast\nabla B_y
		\right),
		\label{eq:FOAM}\\
		\mathbf F_{\rm SAM}
		&= \kappa_{\rm S}/(\mu_0\omega_0)\;
		\big(-\hat{\mathbf z}\times 
		\nabla\,\mathrm{Im}[B_x^\ast B_y]\big).
		\label{eq:FSAM}
	\end{align}
	Here the asterisk denotes complex conjugation, and the coupling coefficients $\kappa_g$, $\kappa_o$, and $\kappa_s$ are determined by the complex magnetic polarizability $\alpha_m$\cite{yang2025optical}.
	
	\begin{figure}[t]
		\includegraphics[width=0.5\textwidth]{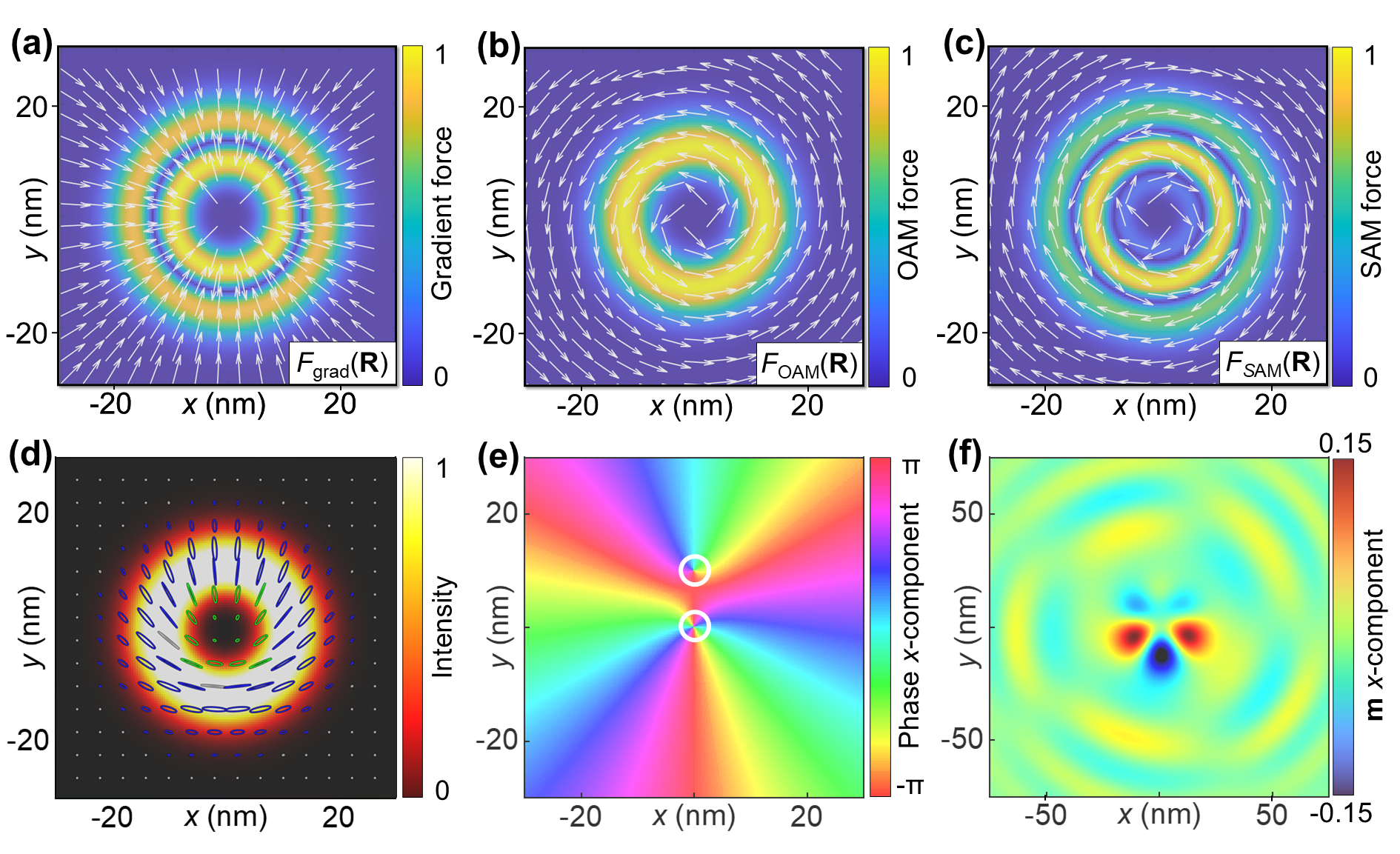}
		\caption{
			(a)–(c) 
			Optical-force distributions showing the gradient, OAM, and SAM components.
			(d) 
			Polarization distribution of the OSk beam.
			(e) 
			Phase distribution of the $B_x$ field.
			(f) 
			Spin-wave radiation excited by the OSk beam in a uniformly magnetized film.
		}
		\label{fig:3}
	\end{figure}

	Normalized maps of these three force components are shown in Figs.~\ref{fig:3}(a)--\ref{fig:3}(c).
	$\mathbf F_{\rm grad}$ establishes the radial confinement profile: its radial component changes sign across the orbit, producing an annular minimum that fixes $R_0$.  
	$\mathbf F_{\rm OAM}$ is azimuthally uniform on the ring (counterclockwise for $l>0$), generating the steady rotational drift.  
	$\mathbf F_{\rm SAM}$ exhibits a narrow sign-reversal band, such that a small radial displacement places the MSk in regions with opposite tangential forces, giving rise to trochoidal modulation.
	The actual orbital is governed by the competition between gyrotropic and dissipative responses in Eq.~(\ref{eq:thiele}), characterized by $|G|/(\alpha D)$:
	small $\alpha$ enhances the phase-delayed Magnus response and strengthens the oscillation, whereas large $\alpha$ suppresses this precessional component and yields smoother motion.
	Together, these forces set the confinement radius, rotation direction, and modulation pattern of the MSk trajectory.

	To elucidate the origin of skipping, we examine the field topology of the OSk.
	As shown in Fig.~\ref{fig:3}(d), the intensity is nearly uniform along the ring, whereas the polarization varies azimuthally, leading to a spatially modulated magnetization--polarization coupling.
	Figure~\ref{fig:3}(e) further reveals spiral wavefronts and localized phase-twist features in the projected $B_x$ component.
	These features originate from the interference between the two LG modes with different OAMs, indicating that the optical drive is intrinsically nonuniform along the orbit.
	The same asymmetry appears in the magnetic response: Fig.~\ref{fig:3}(f) shows that the OSk excites an asymmetric multi-lobe spin-wave pattern even in a uniformly magnetized film without any MSk\cite{fujita2017encoding}, evidencing an azimuthally nonuniform momentum transfer to the magnetic medium.
	By contrast, a single LG mode yields a fully symmetric profile (see Supplemental Material\cite{SM}).

\begin{figure}[t]
		\includegraphics[width=0.5\textwidth]{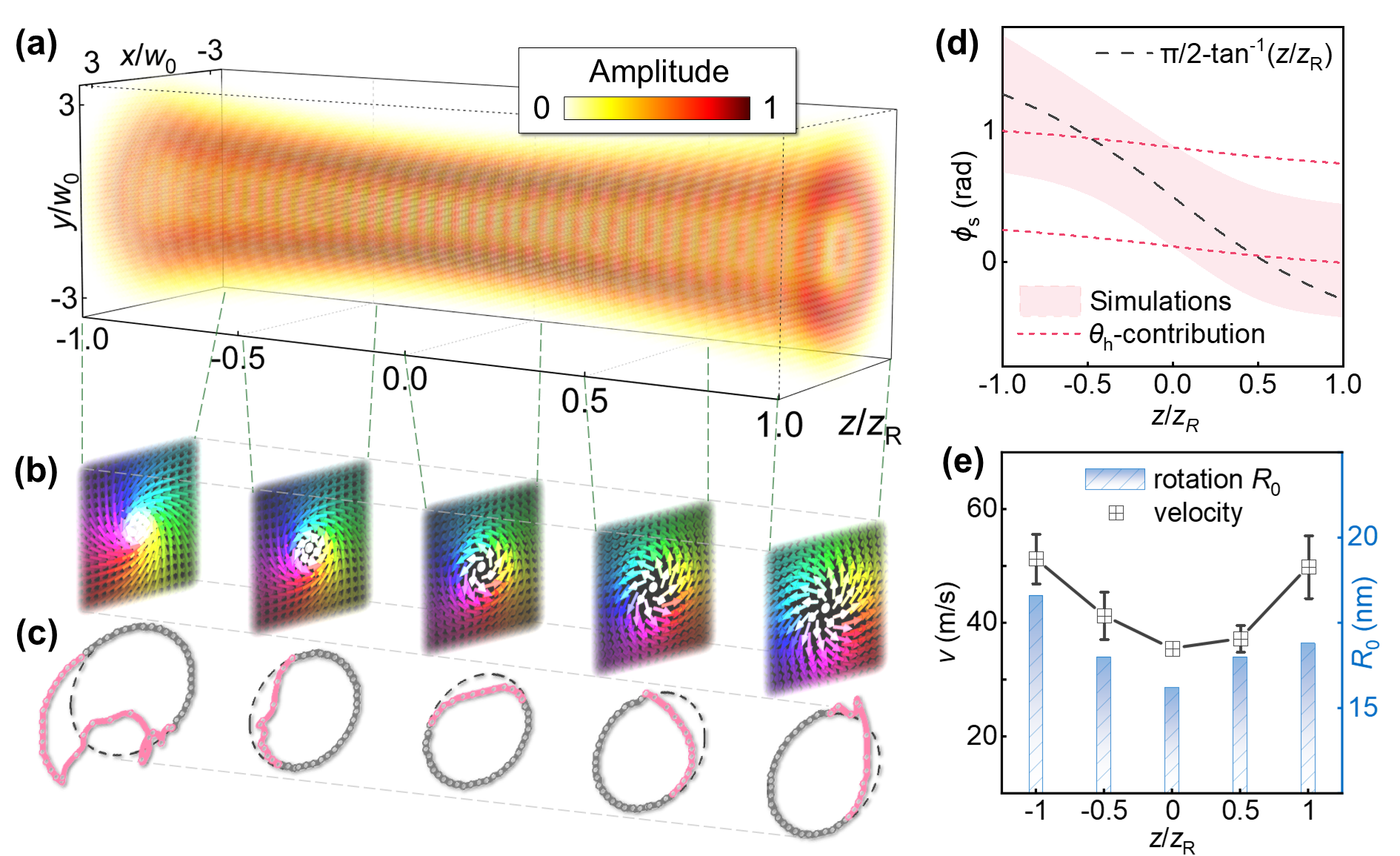}
		\caption{
			(a) 
			3D amplitude distribution of the OSk beam.
			(b), (c) Stokes-vector textures and simulated MSk trajectories at the corresponding cross sections.
			(d) 
			Skipping azimuth $\phi_\mathrm{s}(z)$ from simulations compared with the analytical prediction. 
			(e) 
			Rotation velocity $v$ and orbital radius $R_0$ versus $z/z_\mathrm{R}$.
		}
		\label{fig:4}
	\end{figure}

	\textit{Three-Dimensional Propagation.}--
	We further extend the model to 3D by incorporating the OSk beam propagation along $z$-axis. The 3D LG mode is expressed as
	\begin{equation}
		\begin{aligned}
			u_{n,l}(\rho,\phi,z)
			&= B_0
			\left(\frac{\rho}{W(z)}\right)^{|l|}
			\mathrm{e}^{-\rho^2/W^2(z)}
			L_n^{|l|}\!\left(\frac{2\rho^2}{W^2(z)}\right)\,
			\mathrm{e}^{i l \phi}
			\\
			&\quad\times
			\mathrm{e}^{-\,i\rho^2 z/\,\left[W^2(z)\,z_\mathrm{R}\right]}
			\times
			\mathrm{e}^{-\,i(2n+|l|+1)\,\zeta(z)},
		\end{aligned}
		\label{eq:LG3D}
	\end{equation}
	where $W(z)=W_0\sqrt{1+(z/z_\mathrm{R})^2}$ is the beam waist, $z_\mathrm{R}=\pi W_0^2/\lambda$ is the Rayleigh range, and $\zeta(z)=\tan^{-1}(z/z_\mathrm{R})$ is the Gouy phase.
	Figure~\ref{fig:4}(a) plots the OSk beam amplitude $|\mathbf B_\mathrm{opt}|$, showing a gently confined waist at $z=0$ and a gradual diffraction-induced expansion along $z$-axis.

	In Fig.~\ref{fig:4}(b), the Stokes-vector textures at representative cross sections show that $\theta_\mathrm{h}$ varies by $\pi/4$ within one Rayleigh range.
	Figure~\ref{fig:4}(c) shows the corresponding MSk trajectories, which exhibit a pronounced nonlinear dependence on $z$, with increasing distortions at larger $|z/z_\mathrm{R}|$, particularly for negative $z$.
	This behavior is governed by the differential $\zeta(z)$ of the two LG modes.
	As shown in Eq.~(\ref{eq:LG3D}), each mode carries a Gouy-phase factor $\exp[-i q_j \zeta(z)]$, where $q_j = 2n_j + |l_j| + 1$ ($j=0,1$).
	The resulting relative phase is $\Delta\varPhi(z) = \theta_0 - \Delta q\,\zeta(z)$, with $\Delta q \equiv q_1 - q_0$.
	Since $\zeta(z)=\tan^{-1}(z/z_\mathrm{R})$ is an odd function, $\Delta\varPhi(z)$ reverses sign across the focal plane for any $\Delta q\neq 0$, giving rise to intrinsically nonlinear and $z$-asymmetric trajectories.
	To quantify this effect, Fig.~\ref{fig:4}(d) plots $\phi_\mathrm{s}(z)$ extracted from simulations, together with the $\theta_\mathrm{h}$-contribution and $\zeta(z)$-contribution $\pi/2-\tan^{-1}(z/z_\mathrm{R})$.
	The excellent agreement between the simulations and the analytical curve demonstrates that $\zeta(z)$ dominates the $z$-dependent evolution of the skipping trajectory.
	Near the focal plane ($|z|\ll z_\mathrm{R}$), it can be linearized as $\zeta(z)\simeq z/z_\mathrm{R}$, so that the phase gradient is proportional to $|\Delta q|$. 
	Consequently, $\phi_{\mathrm{s}}$ remains approximately symmetric within $|z/z_\mathrm{R}|\lesssim 0.5$, consistent with Fig.~\ref{fig:4}(c).

	Finally, Fig.~\ref{fig:4}(e) shows the velocity $v$ and the orbit radius $R_0$ as functions of $z/z_\mathrm{R}$. 
	Both quantities vary non-monotonically, reaching minima at $z=0$ and increasing on both sides. 
	This trend reflects the beam expansion and the resulting modulation of the optical force landscape along the propagation direction. 
	Error bars represent the temporal fluctuations of $v$, which also indicate the degree of trajectory deformation.
	
	\textit{Conclusions and outlook.}--
	We have investigated the interaction between a MSk and an OSk beam, identifying three intrinsic dynamical modes: rotation, skipping, and trochoidal motion.
	A force-based analysis reveals that gradient, OAM, and SAM contributions govern radial confinement, azimuthal drift, and precessional modulation, respectively.
	The skipping motion originates from the azimuthal asymmetry of the OSk field and is jointly determined by OSk helicity and MSk chirality.
	Extending the system to 3D introduces a propagation-dependent phase dominated by the Gouy phase, leading to pronounced $z$-asymmetric skipping.
	Our results highlight how the vectorial topology of structured light can engineer emergent magnetic dynamics.
	The three dynamical modes identified here suggest distinct functionalities: rotational confinement may enable tweezer-like trapping and transport of MSk~\cite{kim2025photothermal}, helicity-locked skipping may support phase-selective trajectory steering and magnetic-signal transduction, with possible electrical readout through magnetoresistive detection schemes~\cite{zhao2024electrical}, and trochoidal motion may serve as an optically tunable oscillator-like response~\cite{liu2015dynamical}.
	These features make OSk--MSk coupling promising for contact-free and programmable optomagnonic manipulation.

	\textit{Acknowledgements.}--
	The authors thank Xiaobin Lin for useful discussions. 
	L.B. and M.M. acknowledge support by the Grant-in-Aid for JSPS Fellow (No. JP25KF0118). 
	J.C. acknowledges support by the National Natural Science Foundation of China (No. 12274299) and the Shanghai Pujiang Program (No. 25PJA105). 
	X.Z. and M.M. acknowledge support by the CREST, the Japan Science and Technology Agency (No. JPMJCR20T1). 
	X.Z. also acknowledges support by the Grants-in-Aid for Scientific Research from JSPS KAKENHI (No. JP25K17939 and No.JP20F20363), and support fromthe HKUSTIAS Center for Quantum Matter. 
	M.M. also acknowledges support by the Grants-in-Aid for Scientific Research from JSPS KAKENHI (No. JP25H00611, No. JP24H02231, No. JP23H04522, and No. JP20H00337) and the Waseda University Grant for Special Research Projects (No. 2026C147). 
	Y. Z. acknowledges support by the National Natural Science Foundation of China (No. 12374123), Guangdong Basic Research Center of Excellence for Aggregate Science, and the 2023 SZSTI stable support scheme. 
	C.W.Q. acknowledges support from the Ministry of Education, Republic of Singapore (Grant No. A-8002152-00-00).
	
	\bibliographystyle{apsrev4-2}
	\bibliography{Manuscript}

@book{volovik2003universe,
	title={The universe in a helium droplet},
	author={Volovik, Grigory E},
	volume={117},
	year={2003},
	publisher={OUP Oxford}
}

@article{skyrme1962unified,
	title={A unified field theory of mesons and baryons},
	author={Skyrme, Tony Hilton Royle},
	journal={Nuclear Physics},
	volume={31},
	pages={556--569},
	year={1962},
	publisher={Elsevier}
}

@article{sondhi1993skyrmions,
	title={Skyrmions and the crossover from the integer to fractional quantum Hall effect at small Zeeman energies},
	author={Sondhi, Shivaji Lal and Karlhede, A and Kivelson, SA and Rezayi, EH},
	journal={Physical Review B},
	volume={47},
	number={24},
	pages={16419},
	year={1993},
	publisher={APS}
}

@article{al2001skyrmions,
	title={Skyrmions in a ferromagnetic Bose--Einstein condensate},
	author={Al Khawaja, Usama and Stoof, Henk},
	journal={Nature},
	volume={411},
	number={6840},
	pages={918--920},
	year={2001},
	publisher={Nature Publishing Group UK London}
}

@article{donati2016twist,
	title={Twist of generalized skyrmions and spin vortices in a polariton superfluid},
	author={Donati, Stefano and Dominici, Lorenzo and Dagvadorj, Galbadrakh and Ballarini, Dario and De Giorgi, Milena and Bramati, Alberto and Gigli, Giuseppe and Rubo, Yuri G and Szyma{\'n}ska, Marzena Hanna and Sanvitto, Daniele},
	journal={Proceedings of the National Academy of Sciences},
	volume={113},
	number={52},
	pages={14926--14931},
	year={2016},
	publisher={National Academy of Sciences}
}

@article{muhlbauer2009skyrmion,
	title={Skyrmion lattice in a chiral magnet},
	author={Muhlbauer, Sebastian and Binz, Benedikt and Jonietz, Florian and Pfleiderer, Christian and Rosch, Achim and Neubauer, Anja and Georgii, Robert and Boni, Peter},
	journal={Science},
	volume={323},
	number={5916},
	pages={915--919},
	year={2009},
	publisher={American Association for the Advancement of Science}
}

@article{das2019observation,
	title={Observation of room-temperature polar skyrmions},
	author={Das, S and Tang, YL and Hong, Z and Gon{\c{c}}alves, MAP and McCarter, MR and Klewe, C and Nguyen, KX and G{\'o}mez-Ortiz, F and Shafer, P and Arenholz, E and others},
	journal={Nature},
	volume={568},
	number={7752},
	pages={368--372},
	year={2019},
	publisher={Nature Publishing Group UK London}
}

@article{smalyukh2010three,
	title={Three-dimensional structure and multistable optical switching of triple-twisted particle-like excitations in anisotropic fluids},
	author={Smalyukh, Ivan I and Lansac, Yves and Clark, Noel A and Trivedi, Rahul P},
	journal={Nature Materials},
	volume={9},
	number={2},
	pages={139--145},
	year={2010},
	publisher={Nature Publishing Group UK London}
}

@article{tsesses2018optical,
	title={Optical skyrmion lattice in evanescent electromagnetic fields},
	author={Tsesses, Shai and Ostrovsky, Evgeny and Cohen, Kobi and Gjonaj, Bergin and Lindner, Netanel H and Bartal, Guy},
	journal={Science},
	volume={361},
	number={6406},
	pages={993--996},
	year={2018},
	publisher={American Association for the Advancement of Science}
}

@article{gao2020paraxial,
	title={Paraxial skyrmionic beams},
	author={Gao, Sijia and Speirits, Fiona C and Castellucci, Francesco and Franke-Arnold, Sonja and Barnett, Stephen M and G{\"o}tte, J{\"o}rg B},
	journal={Physical Review A},
	volume={102},
	number={5},
	pages={053513},
	year={2020},
	publisher={APS}
}

@article{ge2021observation,
	title={Observation of acoustic skyrmions},
	author={Ge, Hao and Xu, Xiang-Yuan and Liu, Le and Xu, Rui and Lin, Zhi-Kang and Yu, Si-Yuan and Bao, Ming and Jiang, Jian-Hua and Lu, Ming-Hui and Chen, Yan-Feng},
	journal={Physical Review Letters},
	volume={127},
	number={14},
	pages={144502},
	year={2021},
	publisher={APS}
}

@article{smirnova2024water,
	title={Water-wave vortices and skyrmions},
	author={Smirnova, Daria A and Nori, Franco and Bliokh, Konstantin Y},
	journal={Physical Review Letters},
	volume={132},
	number={5},
	pages={054003},
	year={2024},
	publisher={APS}
}

@article{manton1987geometry,
	title={Geometry of skyrmions},
	author={Manton, NS},
	journal={Communications in Mathematical Physics},
	volume={111},
	number={3},
	pages={469--478},
	year={1987},
	publisher={Springer}
}

@article{wilczek1982quantum,
	title={Quantum mechanics of fractional-spin particles},
	author={Wilczek, Frank},
	journal={Physical Review Letters},
	volume={49},
	number={14},
	pages={957},
	year={1982},
	publisher={APS}
}

@article{jiang2017direct,
	title={Direct observation of the skyrmion Hall effect},
	author={Jiang, Wanjun and Zhang, Xichao and Yu, Guoqiang and Zhang, Wei and Wang, Xiao and Benjamin Jungfleisch, M and Pearson, John E and Cheng, Xuemei and Heinonen, Olle and Wang, Kang L and others},
	journal={Nature Physics},
	volume={13},
	number={2},
	pages={162--169},
	year={2017},
	publisher={Nature Publishing Group UK London}
}

@article{yu2010real,
	title={Real-space observation of a two-dimensional skyrmion crystal},
	author={Yu, XZ and Onose, Yoshinori and Kanazawa, Naoya and Park, Joung Hwan and Han, JH and Matsui, Yoshio and Nagaosa, Naoto and Tokura, Yoshinori},
	journal={Nature},
	volume={465},
	number={7300},
	pages={901--904},
	year={2010},
	publisher={Nature Publishing Group UK London}
}

@article{iwasaki2013current,
	title={Current-induced skyrmion dynamics in constricted geometries},
	author={Iwasaki, Junichi and Mochizuki, Masahito and Nagaosa, Naoto},
	journal={Nature Nanotechnology},
	volume={8},
	number={10},
	pages={742--747},
	year={2013},
	publisher={Nature Publishing Group UK London}
}

@article{sampaio2013nucleation,
	title={Nucleation, stability and current-induced motion of isolated magnetic skyrmions in nanostructures},
	author={Sampaio, Jo{\~a}o and Cros, Vincent and Rohart, Stanislas and Thiaville, Andr{\'e} and Fert, Albert},
	journal={Nature Nanotechnology},
	volume={8},
	number={11},
	pages={839--844},
	year={2013},
	publisher={Nature Publishing Group}
}

@article{fert2013skyrmions,
	title={Skyrmions on the track},
	author={Fert, Albert and Cros, Vincent and Sampaio, Joao},
	journal={Nature Nanotechnology},
	volume={8},
	number={3},
	pages={152--156},
	year={2013},
	publisher={Nature Publishing Group UK London}
}

@article{zhang2015magnetic,
	title={Magnetic skyrmion logic gates: conversion, duplication and merging of skyrmions},
	author={Zhang, Xichao and Ezawa, Motohiko and Zhou, Yan},
	journal={Scientific Reports},
	volume={5},
	number={1},
	pages={1--8},
	year={2015},
	publisher={Nature Publishing Group}
}

@article{zhang2025nanofluidic,
	title={Nanofluidic logic based on chiral skyrmion flows},
	author={Zhang, Xichao and Xia, Jing and Zhou, Yan and Zhao, Guoping and Liu, Xiaoxi and Xu, Yongbing and Mochizuki, Masahito},
	journal={Proceedings of the National Academy of Sciences},
	volume={122},
	number={44},
	pages={e2506204122},
	year={2025},
	publisher={National Academy of Sciences}
}

@article{mochizuki2012spin,
	title={Spin-wave modes and their intense excitation effects in skyrmion crystals},
	author={Mochizuki, Masahito},
	journal={Physical Review Letters},
	volume={108},
	number={1},
	pages={017601},
	year={2012},
	publisher={APS}
}

@article{huang2017magnetic,
	title={Magnetic skyrmion-based synaptic devices},
	author={Huang, Yangqi and Kang, Wang and Zhang, Xichao and Zhou, Yan and Zhao, Weisheng},
	journal={Nanotechnology},
	volume={28},
	number={8},
	pages={08LT02},
	year={2017},
	publisher={IOP Publishing}
}

@article{davis2020ultrafast,
	title={Ultrafast vector imaging of plasmonic skyrmion dynamics with deep subwavelength resolution},
	author={Davis, Timothy J and Janoschka, David and Dreher, Pascal and Frank, Bettina and Meyer zu Heringdorf, Frank-J and Giessen, Harald},
	journal={Science},
	volume={368},
	number={6489},
	pages={eaba6415},
	year={2020},
	publisher={American Association for the Advancement of Science}
}

@article{shen2021supertoroidal,
	title={Supertoroidal light pulses as electromagnetic skyrmions propagating in free space},
	author={Shen, Yijie and Hou, Yaonan and Papasimakis, Nikitas and Zheludev, Nikolay I},
	journal={Nature Communications},
	volume={12},
	number={1},
	pages={5891},
	year={2021},
	publisher={Nature Publishing Group UK London}
}

@article{deng2022observation,
	title={Observation of localized magnetic plasmon skyrmions},
	author={Deng, Zi-Lan and Shi, Tan and Krasnok, Alex and Li, Xiangping and Al{\`u}, Andrea},
	journal={Nature Communications},
	volume={13},
	number={1},
	pages={8},
	year={2022},
	publisher={Nature Publishing Group UK London}
}

@article{du2019deep,
	title={Deep-subwavelength features of photonic skyrmions in a confined electromagnetic field with orbital angular momentum},
	author={Du, Luping and Yang, Aiping and Zayats, Anatoly V and Yuan, Xiaocong},
	journal={Nature Physics},
	volume={15},
	number={7},
	pages={650--654},
	year={2019},
	publisher={Nature Publishing Group UK London}
}

@article{dai2020plasmonic,
	title={Plasmonic topological quasiparticle on the nanometre and femtosecond scales},
	author={Dai, Yanan and Zhou, Zhikang and Ghosh, Atreyie and Mong, Roger SK and Kubo, Atsushi and Huang, Chen-Bin and Petek, Hrvoje},
	journal={Nature},
	volume={588},
	number={7839},
	pages={616--619},
	year={2020},
	publisher={Nature Publishing Group UK London}
}

@article{zhang2021bloch,
	title={Bloch-type photonic skyrmions in optical chiral multilayers},
	author={Zhang, Qiang and Xie, Zhenwei and Du, Luping and Shi, Peng and Yuan, Xiaocong},
	journal={Physical Review Research},
	volume={3},
	number={2},
	pages={023109},
	year={2021},
	publisher={APS}
}

@article{krol2021observation,
	title={Observation of second-order meron polarization textures in optical microcavities},
	author={Kr{\'o}l, Mateusz and Sigurdsson, Helgi and Rechci{\'n}ska, Katarzyna and Oliwa, Przemys{\l}aw and Tyszka, Krzystof and Bardyszewski, Witold and Opala, Andrzej and Matuszewski, Micha{\l} and Morawiak, Przemys{\l}aw and Mazur, Rafa{\l} and others},
	journal={Optica},
	volume={8},
	number={2},
	pages={255--261},
	year={2021},
	publisher={OSA}
}

@article{sugic2021particle,
	title={Particle-like topologies in light},
	author={Sugic, Danica and Droop, Ramon and Otte, Eileen and Ehrmanntraut, Daniel and Nori, Franco and Ruostekoski, Janne and Denz, Cornelia and Dennis, Mark R},
	journal={Nature Communications},
	volume={12},
	number={1},
	pages={6785},
	year={2021},
	publisher={Nature Publishing Group UK London}
}

@article{shen2022generation,
	title={Generation of optical skyrmions with tunable topological textures},
	author={Shen, Yijie and Mart{\'\i}nez, Eduardo Casas and Rosales-Guzm{\'a}n, Carmelo},
	journal={ACS Photonics},
	volume={9},
	number={1},
	pages={296--303},
	year={2022},
	publisher={ACS Publications}
}

@article{yang2023spin,
	title={Spin-manipulated photonic skyrmion-pair for pico-metric displacement sensing},
	author={Yang, Aiping and Lei, Xinrui and Shi, Peng and Meng, Fanfei and Lin, Min and Du, Luping and Yuan, Xiaocong},
	journal={Advanced Science},
	volume={10},
	number={12},
	pages={2205249},
	year={2023},
	publisher={Wiley Online Library}
}

@article{wang2024topological,
	title={Topological structures of energy flow: Poynting vector skyrmions},
	author={Wang, Sicong and Zhou, Zhikai and Zheng, Zecan and Sun, Jialin and Cao, Hongkun and Song, Shichao and Deng, Zi-Lan and Qin, Fei and Cao, Yaoyu and Li, Xiangping},
	journal={Physical Review Letters},
	volume={133},
	number={7},
	pages={073802},
	year={2024},
	publisher={APS}
}

@article{chen2025more,
	title={More than just a name? From magnetic to optical skyrmions and the topology of light},
	author={Chen, Jian and Forbes, Andrew and Qiu, Cheng-Wei},
	journal={Light: Science \& Applications},
	volume={14},
	number={1},
	pages={28},
	year={2025},
	publisher={Nature Publishing Group UK London}
}

@article{fujita2017encoding,
	title={Encoding orbital angular momentum of light in magnets},
	author={Fujita, Hiroyuki and Sato, Masahiro},
	journal={Physical Review B},
	volume={96},
	number={6},
	pages={060407},
	year={2017},
	publisher={APS}
}

@article{yang2018photonic,
	title={Photonic orbital angular momentum transfer and magnetic skyrmion rotation},
	author={Yang, Wenrui and Yang, Huanhuan and Cao, Yunshan and Yan, Peng},
	journal={Optics Express},
	volume={26},
	number={7},
	pages={8778--8790},
	year={2018},
	publisher={Optical Society of America}
}

@article{jiang2020twisted,
	title={Twisted magnon as a magnetic tweezer},
	author={Jiang, Yuanyuan and Yuan, HY and Li, Z-X and Wang, Zhenyu and Zhang, HW and Cao, Yunshan and Yan, Peng},
	journal={Physical Review Letters},
	volume={124},
	number={21},
	pages={217204},
	year={2020},
	publisher={APS}
}

@article{guan2023optically,
	title={Optically controlled ultrafast dynamics of skyrmion in antiferromagnets},
	author={Guan, SH and Liu, Y and Hou, ZP and Chen, DY and Fan, Z and Zeng, M and Lu, XB and Gao, XS and Qin, MH and Liu, J-M},
	journal={Physical Review B},
	volume={107},
	number={21},
	pages={214429},
	year={2023},
	publisher={APS}
}

@article{lei2024rotational,
	title={Rotational motion of skyrmion driven by optical vortex in frustrated magnets},
	author={Lei, YM and Yang, QQ and Tang, ZH and Tian, G and Hou, ZP and Qin, MH},
	journal={Applied Physics Letters},
	volume={125},
	number={7},
	pages={072401},
	year={2024},
	publisher={AIP Publishing}
}

@article{zhang2026skyrmion,
	title={Skyrmion generation through the chirality interplay of light and magnetism},
	author={Zhang, Qifan and Lin, Shirong and Zhang, Wu},
	journal={Communications Physics},
	volume={9},
	number={55},
	pages={55},
	year={2026},
	publisher={Nature Publishing Group UK London}
}

@article{fujita2017ultrafast,
	title={Ultrafast generation of skyrmionic defects with vortex beams: Printing laser profiles on magnets},
	author={Fujita, Hiroyuki and Sato, Masahiro},
	journal={Physical Review B},
	volume={95},
	number={5},
	pages={054421},
	year={2017},
	publisher={APS}
}

@incollection{evans2013mechanical,
	title={Mechanical astronomy: A route to the ancient discovery of epicycles and eccentrics},
	author={Evans, James and Cannan, Christi{\'a}n Carlos},
	booktitle={From Alexandria, through Baghdad: Surveys and studies in the ancient Greek and medieval Islamic mathematical sciences in honor of JL Berggren},
	pages={145--174},
	year={2013},
	publisher={Springer}
}

@article{ritzmann2018trochoidal,
	title={Trochoidal motion and pair generation in skyrmion and antiskyrmion dynamics under spin--orbit torques},
	author={Ritzmann, Ulrike and von Malottki, Stephan and Kim, Joo-Von and Heinze, Stefan and Sinova, Jairo and Dup{\'e}, Bertrand},
	journal={Nature Electronics},
	volume={1},
	number={8},
	pages={451--457},
	year={2018},
	publisher={Nature Publishing Group UK London}
}

@article{shen2024optical,
	title={Optical skyrmions and other topological quasiparticles of light},
	author={Shen, Yijie and Zhang, Qiang and Shi, Peng and Du, Luping and Yuan, Xiaocong and Zayats, Anatoly V},
	journal={Nature Photonics},
	volume={18},
	number={1},
	pages={15--25},
	year={2024},
	publisher={Nature Publishing Group UK London}
}

@article{wang2025perturbation,
	title={Perturbation-resilient integer arithmetic using optical skyrmions},
	author={Wang, An Aloysius and Ma, Yifei and Zhang, Yunqi and Zhao, Zimo and Cai, Yuxi and Qiu, Xuke and Dong, Bowei and He, Chao},
	journal={Nature Photonics},
	volume={19},
	pages={1367--1375},
	year={2025}
}

@article{ma2025nanophotonic,
	title={Nanophotonic quantum skyrmions enabled by semiconductor cavity quantum electrodynamics},
	author={Ma, Jiantao and Yang, Jiawei and Liu, Shunfa and Chen, Bo and Li, Xueshi and Song, Changkun and Qiu, Guixin and Zou, Kai and Hu, Xiaolong and Li, Feng and Yu, Ying and Liu, Jin},
	journal={Nature Physics},
	volume={21},
	number={9},
	pages={1462--1468},
	year={2025},
	publisher={Nature Publishing Group UK London}
}

@article{yang2025optical,
	title={Optical sorting: past, present and future},
	author={Yang, Meng and Shi, Yuzhi and Song, Qinghua and Wei, Zeyong and Dun, Xiong and Wang, Zhiming and Wang, Zhanshan and Qiu, Cheng-Wei and Zhang, Hui and Cheng, Xinbin},
	journal={Light: Science \& Applications},
	volume={14},
	number={1},
	pages={103},
	year={2025},
	publisher={Nature Publishing Group UK London}
}

@article{zhang2021autonomous,
	title={Autonomous materials systems from active liquid crystals},
	author={Zhang, Rui and Mozaffari, Ali and de Pablo, Juan J},
	journal={Nature Reviews Materials},
	volume={6},
	number={5},
	pages={437--453},
	year={2021},
	publisher={Nature Publishing Group UK London}
}

@article{asilehan2025light,
	title={Light-driven dancing of nematic colloids in fractional skyrmions and bimerons},
	author={Asilehan, Zhawure and Tang, Wentao and Zhang, Jing and Chen, Zijun and Wang, Ruijie and Shi, Qingtian and Song, Ganlin and Jiang, Jinghua and Zhang, Rui and Peng, Chenhui},
	journal={Nature Communications},
	volume={16},
	number={1},
	pages={1148},
	year={2025},
	publisher={Nature Publishing Group UK London}
}

@article{asilehan2025monopole,
	title={Monopole-mediated light control of half skyrmion topology in nematic liquid crystals},
	author={Asilehan, Zhawure and Tang, Wentao and Zheng, Xinda and Wang, Ruijie and Zhang, Jing and Tian, Kun and Vergara, Fernando and Shi, Qingtian and Chen, Zijun and Jiang, Jinghua and others},
	journal={Nature Communications},
	volume={16},
	number={1},
	pages={9178},
	year={2025},
	publisher={Nature Publishing Group UK London}
}

@article{kim2025photothermal,
	title={Photothermal skyrmion tweezer: programmable optical manipulation of magnetic topological quasiparticles},
	author={Kim, Jaeyu and Yang, Seungmo and Kim, Dongha and Moon, Kyoung-Woong and Kim, Changsoo and Hwang, Chanyong and Seo, Min-Kyo},
	journal={Nature Communications},
	volume={16},
	pages={11375},
	year={2025},
	publisher={Nature Publishing Group UK London}
}

@article{zhao2024electrical,
	title={Electrical detection of mobile skyrmions with 100\% tunneling magnetoresistance in a racetrack-like device},
	author={Zhao, Mengqi and Chen, Aitian and Huang, Pei-Yuan and Liu, Chao and Shen, Lisen and Liu, Jia and Zhao, Lian and Fang, Baisheng and Yue, Wen-Chao and Zheng, Dong and Wang, Lei and Bai, Haoliang and Shen, Kai and Zhou, Yan and Wang, Shouguo and Liu, Enlong and He, Shoujie and Wang, Ya-Li and Zhang, Xichao and Jiang, Wanjun},
	journal={npj Quantum Materials},
	volume={9},
	pages={50},
	year={2024},
	publisher={Nature Publishing Group UK London}
}

@article{liu2015dynamical,
	title={Dynamical skyrmion state in a spin current nano-oscillator with perpendicular magnetic anisotropy},
	author={Liu, R. H. and Lim, W. L. and Urazhdin, S.},
	journal={Physical Review Letters},
	volume={114},
	pages={137201},
	year={2015},
	publisher={APS}
}

@misc{SM,
	note = {See Supplemental Material at [URL will be inserted by publisher] for more
	details, which includes Refs.~\cite{vansteenkiste2014design,heeres2014subwavelength}.}
}

@article{vansteenkiste2014design,
	title={The design and verification of MuMax3},
	author={Vansteenkiste, Arne and Leliaert, Jonathan and Dvornik, Mykola and Garcia-Sanchez, Felipe and Van Waeyenberge, Bartel},
	journal={AIP Advances},
	volume={4},
	pages={107133},
	year={2014},
	publisher={AIP Publishing}
}

@article{heeres2014subwavelength,
	title={Subwavelength focusing of light with orbital angular momentum},
	author={Heeres, Reinier W. and Zwiller, Valery},
	journal={Nano Letters},
	volume={14},
	pages={4598--4601},
	year={2014},
	publisher={American Chemical Society}
}

\end{document}